\documentclass[preprint
]{revtex4}
\newcommand{\be}{\begin{equation}}
\newcommand{\ee}{\end{equation}}
\newcommand{\bea}{\begin{eqnarray}}
\newcommand{\eea}{\end{eqnarray}}
\usepackage{graphicx}
\usepackage{epsfig}
\usepackage{bm}

\usepackage{amssymb}

\begin{document}
\title{Markov Chain Modeling of Polymer Translocation Through Pores}
\author{Felipe Mondaini}
\author{L. Moriconi\footnote{Email of corresponding author: moriconi@if.ufrj.br}}
\affiliation{Instituto de F\'\i sica, Universidade Federal do Rio de Janeiro, \\
C.P. 68528, 21945-970, Rio de Janeiro, RJ, Brazil}
\begin{abstract}
We solve the Chapman-Kolmogorov equation and study the exact splitting probabilities of the general stochastic process which describes polymer translocation through membrane pores within the broad class of Markov chains. Transition probabilities which satisfy a specific balance constraint provide
a refinement of the Chuang-Kantor-Kardar relaxation picture of translocation, allowing us to investigate finite size effects in the evaluation of dynamical scaling exponents. We find that (i) previous Langevin simulation results can be recovered only if corrections to the polymer mobility exponent are taken into account and that (ii) the dynamical scaling exponents have a slow approach to their predicted asymptotic values as the polymer's length increases. We also address, along with strong support from additional numerical simulations, a critical discussion which points in a clear way the viability of the Markov chain approach put forward in this work.
\end{abstract}
\maketitle

\section{Introduction}

The phenomenon of polymer translocation through membrane pores has received a great deal of attention in recent years \cite{muthu}. Important issues are related to the translocation of complex biomolecules in the methabolism of living cells \cite{nelson}, and, on the technological forefront, to developments in the fields of targeted drug/gene delivery \cite{zanta} and DNA sequencing \cite{kasia,fologea,luo}.

It has been known, however, that there is not a universal mechanism for polymer translocation. A case-by-case analysis is necessary to conclude if translocation is related to specific features of the biochemical environment which surrounds the membrane or to the existence of biomolecular motors, as the ones found in mitochondria \cite{dolezal}. Taking a look at the voluminous literature on the subject, one finds studies of polymer translocation driven by chemical or electric potential gradients \cite{kasia, grosberg, huopa, kaifu, wei,luo2},  chaperone-assisted rectified brownian motion \cite{meller} or simply unbiased translocation \cite{wei, luo2,gauthier, panja}.

The case of unbiased translocation, where a polymer is let to diffuse through a membrane pore as the sole consequence of thermal fluctuations, is the ideal starting point for the investigation of more sophisticated models. Chuang, Kantor and Kardar (CKK) \cite{kardar} have introduced a successful description of unbiased homopolymer translocation, assuming, essentially, that the polymer's evolution does not take it far from its equilibrium states. Astonishingly simple as it may sound, translocation is, then, ruled by the diffusion exponent of the polymer center of mass, as if there were no blocking membrane.

We are interested to provide a general kinetic description of polymer translocation and, in particular, of the CKK scaling results. Elementary stochastic events of the underlying Markov chain processes are given by the translocation of individual monomers. As we show in this work, the CKK dependence of the translocation time with polymer size is indeed recovered by the Markov chain modeling, after finite-size corrections are properly eliminated -- an issue not addressed by the original CKK approach.

This paper is organized as follows. In sec. II, we write down the general Chapman-Kolmogorov equation which describes polymer translocation as a Markov chain, and discuss its exact asymptotic solution for the probability of complete translocation. The CKK picture is then taken into account as a way to devise expressions for the transition probabilities of individual monomer translocation events. The main point in our work is that the CKK scaling exponents may be considerably affected by finite size effects, which are partially encoded in corrections to the polymer mobility exponent. 

In sec. III, we compare results obtained from the numerical solution of the Chapman-Kolmogorov equation with the ones from previous detailed Langevin simulations \cite{wei}.

As polymer translocation has been conjectured to be subdiffusive \cite{chatelain} and non-markovian \cite{panja2,dubbeldam}, we address, in sec. IV, a critical analysis of these issues, finding support, from further Langevin simulations, for the pertinence of the Markovian framework. In section V, we summarize our results and point out directions of further research.

\section{Chapman-Kolmogorov Approach}

Our essential aim is to model polymer translocation as a discrete Markov stochastic process. Assume, for a homopolymer of length $N$, that at an arbitrary time instant there are $n$ monomers on the trans side of the membrane and $N-n$ monomers on the cis side, as depicted in Fig. 1. Let $p_n$ and $q_n$ be the probabilities that correspond to cis $\rightarrow$ trans and trans $\rightarrow$ cis monomer transitions, respectively. It is not necessary to have $p_n + q_n =1$; actually, the polymer may get momentarily stuck around the membrane with probability $1 - p_n- q_n$.

\begin{figure}[tbph]
\includegraphics[width=11.88cm, height=9.24cm]{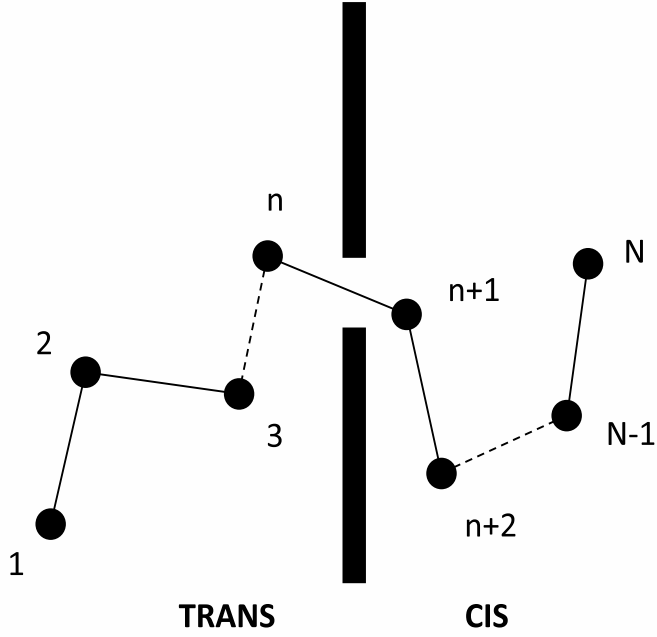}
\caption{The trans and cis  sides of the membrane have, respectively, $n$ and $N-n$ monomers.}
\label{fig1}
\end{figure}

An object of central interest is the probability $P(n,t)$ of finding $n$ monomers on the trans side of the membrane at time $t$, once the initial probability vector $P(n,0)$ is given ($t$ is an integer valued variable). In numerical experiments, it is usual to have $P(n,0) = \delta(n,N/2)$, meaning that the initial state has half of the monomers on each side of membrane. The Chapman-Kolmogorov equation for the translocation process can be readily written as
\bea
&&P(n,t+1) = q_{n+1} P(n+1,t)+p_{n-1} P(n-1,t) \nonumber \\
&&+(1-p_n-q_n)P(n,t) \ , \
\label{c-k}
\eea
where we take $0 \leq n \leq N$, with $p_{-1}=p_0=q_0=q_{N+1}=p_N=q_N=0$, so that translocation proceeds until $n=0$ or $n=N$
($n=0$ and $n=N$ are absorbing states of the stochastic process).

The important task now is to find expressions for the transition probabilities $p_n$ and $q_n$. In the CKK description of translocation \cite{kardar}, it is assumed that the translocation time $\tau_N$ of a polymer of size $N$ is of the order of the Rouse relaxation time \cite{rouse}, that is, the time spent by the center of mass of a free polymer to diffuse along its own gyration radius $R_G \sim N^\nu$. Writing the difusion constant of the center of mass as $\bar D =D/N^\delta$, where $D$ is the diffusion constant of a single monomer, we will have $R_G^2 \sim \bar D  \tau$, and, thus
\be
\tau_N \sim N^{\delta+2 \nu} \ . \
\label{tau}
\ee
The CKK expression $\tau_N \sim N^ {1 + 2 \nu}$ follows by the substitution of $\delta =1$ (as it applies for free polymers in the absence of hydrodynamical couplings) in (\ref{tau}). In our discussion, however, we take $\delta$ to be an adjustable parameter, which may slightly depart from unit for small chains due to pore/membrane interactions. An estimate of $\delta$ can be obtained by the use of the polymer generalization of the usual Einstein relation which establishes the proportionality between the mobility and the diffusion coefficients in brownian motion \cite{DeGennes}. 
We recall that while the mobility can be computed from the velocity response of the polymer to external forces, the diffusion coefficient refers to the polymer evolution in the absence of any external perturbations. In this respect, we point out that recent Langevin simulations \cite{bhatta} indicate that the mobility computed for driven translocation scales as $N^{-\delta}$, with $\delta = 0.81 \pm 0.04$ for polymer sizes $N\leq 256$.

It is important to note, before proceeding, that the original master equation approach introduced by Muthukumar \cite{muthu2}, which is based on a relaxation-to-equilibrium formalism, yields, very generally, the scaling law $\tau \sim N^2$ for translocation. This result is a consequence from the fact that the transition probability ratio $q_n/p_n$ would be given, according to the detailed-balance hypothesis of Ref. \cite{muthu2}, by
\be
\frac{q_n}{p_n}= \exp \left \{(\gamma -1) \left [\frac{1}{n}-\frac{1}{N-n} \right ] \right \} \ , \
\ee 
which is close to unit for $1 < n < N$ (above, $0.5 \leq \gamma \leq 1$ is the scaling exponent used in the definition of the partition function, $Z_n \sim n^{\gamma -1}$, for a polymer chain anchored at one endpoint to a hard wall \cite{muthu2,eisen}). The challenge we are faced, thus, is how to obtain the anomalous scaling predicted by the CKK picture from a markovian line of argumentation. A kinematical solution of this problem (i.e., not based on free-energy considerations) will be provided by an alternative choice of the transition probabilities $p_n$ and $q_n$. As a bonus, we will be able to address the role of finite size effects in the translocation process.

For a heuristic derivation of $p_n$ and $q_n$, assume that the time lapses for single trans $\rightarrow$ cis and cis $\rightarrow$ trans monomer translocations are proportional to $\tau_n /n$ and $\tau_{N-n}/(N-n)$, respectively. This implies that the transition probabilities satisfy the
balance constraint,
\be
q_n \frac{\tau_n}{n}= p_n \frac{\tau_{N-n}}{N-n} = {\hbox{constant}} \ . \  \label{omega}
\ee
Substituting (\ref{tau}) in (\ref{omega}), it follows that
\bea
&&p_n = \frac{c}{(N-n)^{\delta+2 \nu - 1}} \ , \ \nonumber \\
&&q_n = \frac{c} { n^{\delta+2 \nu - 1}} \ , \
\label{prob}
\eea
where $0< c < 1$ is an arbitrary constant. As we will see in the next section, solutions of the Chapman-Kolmogorov equation (\ref{c-k}) with transition probabilities (\ref{prob}) will lead to translocation times which are not given by (\ref{tau}), due to the existence of finite-size effects, a main point of attention in our work. We find that (\ref{tau}) is recovered only in the asymptotic limit of large polymer chains.

The probability of complete translocation to the trans side of the membrane, $P(N,\infty)$, known as the ``splitting probability" in the theory of stochastic process, is available in closed analytical form \cite{vanKampen}. Since this quantity depends on the initial number of trans monomers, $n$, we write $P(N,\infty) \equiv P(n)$, with

\be
P(n) = \frac{1 + \sum_{i=1}^{n-1} \prod_{j=1}^i \frac{q_j}{p_j}}{1 + \sum_{i=1}^{N-1} \prod_{j=1}^i \frac{q_j}{p_j}} \ . \
\label{pn}
\ee
We have performed Langevin simulations in order to check if the above exact splitting probabilities are in fact pertinent. The results, which support the viability of the Markov chain approach to polymer translocation, are reported in sec. IV.

\section{Scaling Regimes}

The main difficulty in carrying out realistic Langevin numerical simulations of polymer translocation is usually related to limitations in the polymer sizes (typically, present computer desktop resources allow one to work finely with a few hundred monomers). It turns out that scaling results, like Eq. (\ref{tau}), are unavoidably affected by finite size scaling effects, so that even if statistical error bars are taken into account, predictions and observations may not satisfactorily match.

We address in the following a comparison between Markov chain modeling results and the ones obtained through the extensive Langevin simulations reported in Ref. \cite{wei}. The general strategy is to numerically solve the Chapman-Kolmogorov equation for a given set of transition probabilities $p_n$, $q_n$ (corresponding to different translocation regimes) and several polymer sizes $N$, always taking the initial probability vector as $P(n,0) = \delta(n,N/2)$.

Let $r$ be the translocation time for an individual realization of polymer translocation.  It is clear that
\be
r=\sum_{s=1}^\infty [1-\Theta(s-r)] \ , \
\ee
where $\Theta(s-r)$ is the Heaviside step function of $s-r$ (we are using the convention $\Theta(0) = 0$), so that the mean translocation 
time can be written simply as
\bea
\tau &=& \langle r \rangle = \sum_{s=1}^\infty [1-\langle \Theta(s-r) \rangle] \nonumber \\
&=& \sum_{s=1}^\infty [1-P(N,s)-P(0,s)] \ , \ \label{taub}
\eea
where the above averages are taken over the ensemble of polymer translocation realizations. In (\ref{taub}), $P(N,s)+P(0,s)$ is the probability, 
to be obtained from the numerical solution of the Chapman-Kolmogorov equation, for the occurrence of complete translocation up to time $s$; it is a monotonically increasing function of $s$, which approaches unit for $s \rightarrow \infty$. For the sake of clarity, note that if $F_s$ is the probability that complete translocation takes place for the first time at time $s$, then $P(N,s)+P(0,s) = \sum_{s'=0}^s F_{s'}$ (recall that translocation is modeled here as a stochastic process with absorbing boundary conditions). Since $\sum_{s'=0}^\infty F_{s'} = 1$, we get $1 - P(N,s)-P(0,s) = \sum_{s'=s+1}^\infty F_{s'}$, and, therefore,
from (\ref{taub}),
\be
\tau = \sum_{s=0}^\infty \sum_{s'=s+1}^\infty F_{s'} = \sum_{s=1}^\infty sF_s \ , \
\ee
which is the more familiar (but unpractical, for our purposes) way of writing $\tau$ as a mean first passage time.

It suffices, for excellent numerical convergence, to retain in (\ref{taub}) all the contributions which have $P(N,s) + P(0,s) < 1 - 10^{-6}$ (results have precision better than $10^{-3} \%$). The scaling exponent $\alpha$ in $\tau \sim N^\alpha$ is then determined by straightforward linear regression on log-log plots.

We assume, in the quasi-equilibrium regimes discussed here, that both sides of the membrane are characterized by the same coil-globule transition temperature $T_\theta$, i.e., in a real experiment the solvent qualities would be the same on both sides of the diathermic membrane. There are, roughly, three possible translocation regimes depending on the equilibrium temperature $T$: (i) $T < T_\theta$, (ii) $T= T_\theta$ and (iii) $T > T_\theta$. We model these regimes, respectively, by gyration radius exponents \cite{DeGennes,havlin,guillou} (i) $\nu = 1/3$ (globule phase), (ii) 
$\nu = 1/2$ ($\theta$-point), and (iii) $\nu = 0.588$ (self avoiding walk).

We have found through numerical solutions of the Chapman-Kolmogorov equation that the translocation time scales with the polymer size in all of the above situations. In order to compare our results with the ones from Langevin simulations \cite{wei} we have limited the number of monomer sizes up to $N=500$. The scaling exponents and translocation time profiles for different translocation regimes are shown in Table I and Fig. 2, respectively. The finite size corrections are found to be in good agreement with the Langevin simulation results if one takes $\delta \simeq 0.88 \pm 0.03$. We note that the discrepancy for the $T=T_\theta$ regime is probably due to lack of precision in locating the $\theta$-point in Langevin simulations (no error bars are reported in Ref. \cite{wei} for $T=T_\theta$). In passing, we call attention to the fact that as $T$ drops much below $T_\theta$, it has been reported, in the same referred work, that the dynamical exponent $\alpha_L$ (defined in Table I) stops decreasing and eventually grows to values larger than $2$. It is possible that such a non-monotonic behavior of $\alpha_L$ has to do with known deviations of Rouse's theory at low temperatures \cite{milchev}.
\vspace{0.2cm}

\begin{figure}[tbph]
\includegraphics[width=11.88cm, height=9.24cm]{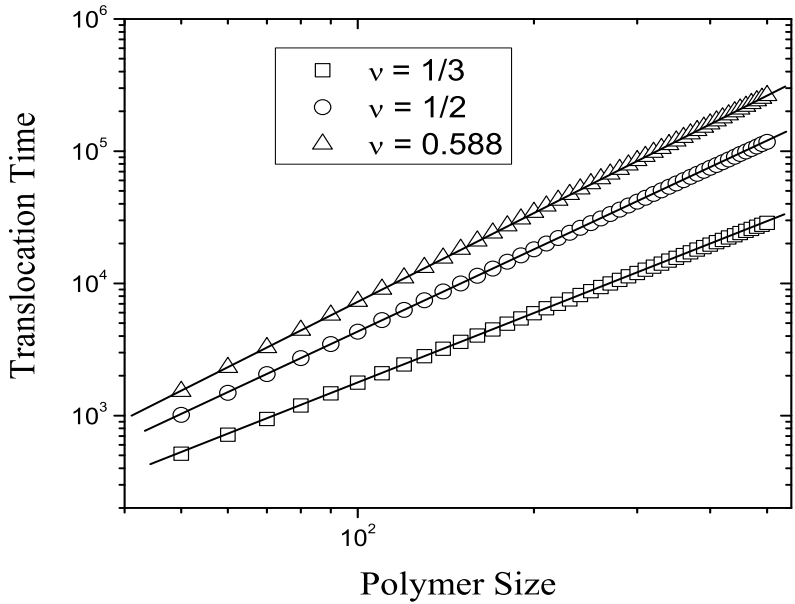}
\caption{The translocation time $\tau$ is found to scale with the polymer's size $N$ as $\tau \sim N^\alpha$. 
We report scaling profiles for $50 \leq N \leq 500$ and $T < T_\theta$ (squares), $\alpha = 1.739(2)$; 
$T = T_\theta$ (circles), $\alpha =2.059(1)$; $T > T_{\theta_{trans}}$ (triangles), $\alpha =2.231(1)$.
We have considered $\delta = 0.88$.}
\label{fig2}
\end{figure}

\begin{table}[tbph]
\begin{tabular}{|c|c|c|c|} \hline
~ Temperatures ~&~ $\alpha_{th}$ ~ & ~ $\alpha_L$  ~ & ~ $\alpha_M $  \\ \hline
$T < T_\theta$  & $5/3$ & $1.74(3)$ & 1.75(3) \\ \hline
$T = T_\theta$  & $2$ & $2.2(?)$ & 2.07(3) \\ \hline
$T > T_\theta$  & $ 2.176 $ & $2.23(3)$ & 2.24(3) \\ \hline
\end{tabular}
\caption{Comparison between scaling exponents.
Here, we assume $\alpha_{th} = 1+ 2\nu$ \cite{kardar} for the quasi-equilibrium cases, while $\alpha_{L}$ and $\alpha_M$ are determined through Langevin simulations \cite{wei} and the Markov chain approach (with $\delta = 0.88 \pm 0.03$). Both $\alpha_L$ and $\alpha_M$ are evaluated 
for polymer sizes in the range $50 \leq N \leq 300$. The error bar for the Langevin simulation with $T=T_\theta$ is not available.}
\end{table}

The crucial point here is that due to finite size effects, the scaling exponents evaluated from Langevin simulations could seem to contradict the theoretical CKK predictions, even if error bars are taken into account in numerical evaluations. The translocation time exponent for a polymer of size $N$ can be  written, in general, as
\be
\alpha = \delta(N) + 2 \nu + f(\delta(N) + 2 \nu,N) \ , \
\ee
with $\delta(N) <1$, $f(\delta(N) + 2 \nu,N) > 0$, and 
\bea
&&\lim_{N \rightarrow \infty} \delta(N) =1 \ , \ \nonumber \\
&&\lim_{N \rightarrow \infty} f(\delta(N) + 2 \nu,N) = 0 \ . \  \label{fse}
\eea
We expect that $\delta(N)$ and $f(\delta(N) + 2 \nu, N)$ are relatively slow functions of $N$. 

For a fixed value of $\delta$, we may compare evaluations of the scaling exponents obtained from the Markov chain approach for progressively larger polymers, with those proposed within the CKK framework. The finite size scaling results are shown in Fig. 3, where we take $\delta =1$ just for the
sake of illustration. One finds, in fact, convergence towards the conjectured values of the scaling exponents.

\begin{figure}[tbph]
\includegraphics[width=11.88cm, height=9.24cm]{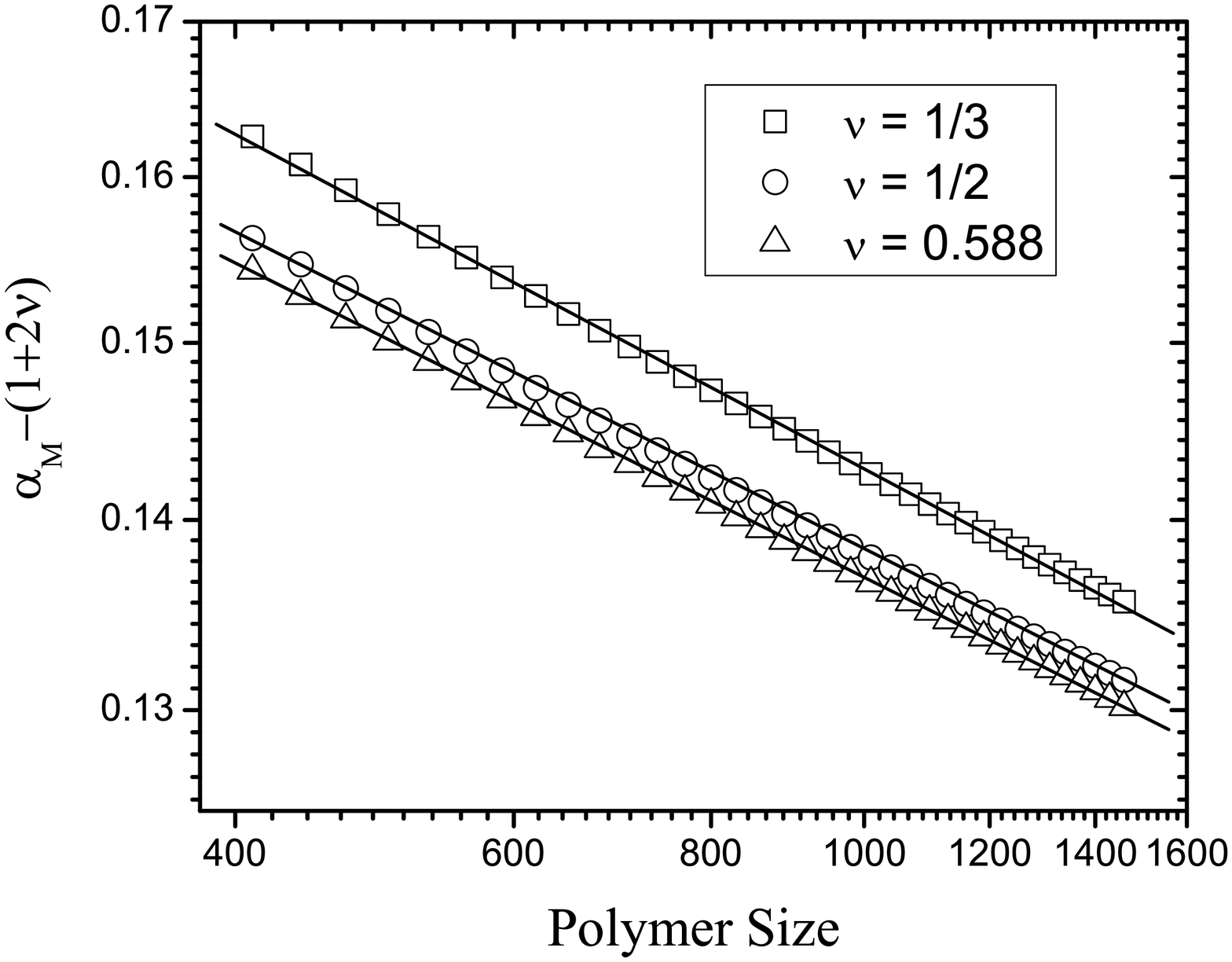}
\caption{Finite size scaling behavior of $\alpha-(1+ 2\nu)$ (we have taken $\delta=1$). Corrections to the scaling exponents $\alpha_{th} = 1 + 2\nu$ 
vanish as a power law $N^{-\beta}$ with scaling exponents $\beta = 0.1418(5)$ (squares), $\beta = 0.1348(4)$ (circles), and 
$\beta = 0.1327(3)$ (triangles).}
\label{fig3}
\end{figure}

Our data is consistent with $\alpha - \alpha_{th} \sim N^{-\beta}$, where $\beta \simeq 0.13 - 0.14$. This scaling behavior implies that the translocation time can be written, taking into account subleading corrections, as
\be
\tau \sim N^{1+2 \nu}[1 + c N^{-\beta} \log (N)] \ . \
\ee
The theoretical computation of $\beta$ is an important challenge deserved for further studies. It is worth emphasizing that a complete discussion on the universality of the translocation exponents should necessarily come along with finite size scaling arguments.

\section{Evidence of Markovian Behavior}

The Markov approach of this work stands in contrast to previous 
claims that memory effects are relevant in the phenomenon of polymer 
translocation. We intend here to go deeper in this issue, first through a critical 
account of the non-markovian line of thought, and then by pointing out further 
results which provide clear support for a markovian description of polymer 
translocation.
 
Let $s(t)$ be the number of monomers on a given side of the membrane at time $t$, in 
the course of unbiased polymer translocation. In the wake of the CKK's picture of 
translocation, it has been conjectured that $s(t)$ should be modeled as a non-markovian 
process. The chain of ideas which would suggest the non-markovian character of polymer 
translocation can be summarized as follows:
\vspace{0.1cm}

(i) Defining the time-dependent variance of the translocation coordinate $s(t)$ as 
$\Delta(t) \equiv \langle s(t)^2 \rangle - \langle s(t) \rangle^2$, one
expects that $\Delta(\tau) \sim N^2$ holds for a polymer of size $N$,
where $\tau$ is the translocation time. This result would naturally follow 
from the anomalous diffusion law $\Delta(t) \sim t^\frac{2}{1 + 2 \nu}$;

(ii) Since $\nu > 0.5$ in the polymer high-temperature phase, 
the anomalous diffusion exponent $2/(1 + 2 \nu) < 1$ indicates that polymer 
translocation is a non-markovian process;

(iii) Monte Carlo simulations of polymer translocation, as performed within the framework
of the Bond Fluctuation Model (BFM) \cite{dubbeldam_bfm}, are able to reproduce the anomalous diffusive profile 
advanced in (i).
\vspace{0.1cm}

It is important to emphasize that there is a large room for questioning the relevance 
of the above statements, as we point out below, in respective order.
\vspace{0.1cm}

(i') There is no universal relation between the dynamic exponents associated to the translocation 
time $\tau$ and the diffusion law. They are completely independent as a matter of principle. 
Actually, we have checked from solutions of the Chapman-Kolmogorov equation (\ref{c-k}) with the transition probabilities 
(\ref{prob}) (which lead, as discussed in the previous sections, to the CKK translocation exponent $1 + 2 \nu$ 
in the limit  of large polymers) that most of the translocation process is described by the normal diffusion law, 
where $\Delta(t)$ is a linear function of time;

(ii') Anomalous diffusion is by no means a sufficient condition for non-markovian behavior. 
There are abundant and important examples in the literature of either super or subdiffusive 
markovian processes \cite{ott_etal}. We recall the theory of Levy flights, just to quote a 
celebrated instance of Markov processes which have anomalous diffusion exponents \cite{mandelbrot};

(iii') Even though Monte Carlo algorithms are not able in general to address dynamical aspects 
of statistical systems, the BFM has been accomplished as a very useful tool for the study of dynamical 
phenomena in polymer physics \cite{paul_etal}. It is clear that the all the credit for the BFM's 
approach relies on the comparison of its predictions with the outcomes of real and numerical experiments. 
Despite the popularity of the BFM strategy in polymer physics, it is not obvious a priori if under 
the specific boundary conditions related to polymer translocation, BFM will provide physically 
meaningful results \cite{panja-barkema}.
\vspace{0.1cm}

Regarding (iii'), in particular, our attention is drawn to recent extensive Langevin simulations of unbiased 
polymer translocation which have been performed in connection with the problem of anomalous diffusion \cite{dubbeldam}. 
Its is claimed, in that work, that on the basis of standard log-log plots, anomalous diffusion is verified
with dynamic exponent $2/(1 + 2 \nu)$ at intermediate time scales. However, we have found that if the very same data 
is represented in linear scales, as in Fig. 4, an excellent and unique linear fit for $\Delta(t)$ holds both for small 
and intermediate time scales. The Langevin simulations of Ref. \cite{dubbeldam} seem to give support, actually, to normal 
diffusion in the form $\Delta(t) = a + bt$ (we note that it is a straighforward exercise to derive arbitrary $a$ and $b$ 
coefficients from a simple collored gaussian noise version of the Ornstein-Uhlenbeck process, which is essentially
markovian for time scales larger than the noise correlation time).

\begin{figure}[tbph]
\includegraphics[width=11.88cm, height=9.24cm]{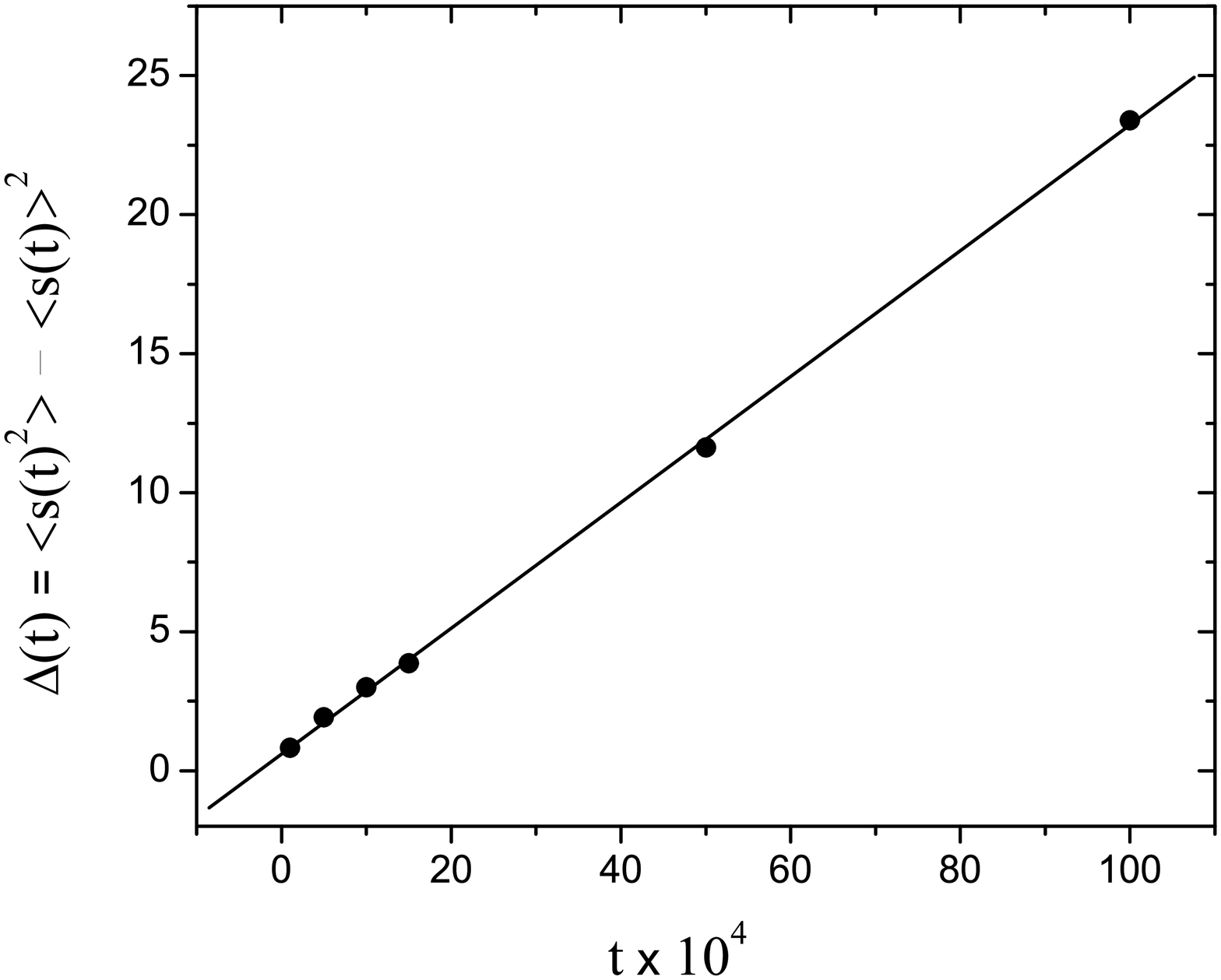}
\caption{The numerical data listed in Fig. 4a of Ref. \cite{dubbeldam} is alternatively plotted in linear scales. 
The linear fit strongly suggests that for these time scales we have $\Delta(t) = a + bt$, with $a=0.62$ and 
$b= 2.3 \times 10^{-5}$.}
\label{fig4}
\end{figure}

We have also carried out Langevin simulations in order to investigate the splitting probabilities predicted by Eq. (\ref{pn}).
While we have not produced a large number of realizations (compared to the 5000 of Ref. \cite{dubbeldam}) which would allow us
to compute the diffusion exponent with reasonable precision, we have collected 560 complete translocation processes, which are
enough to address a numerical test of Eq. (\ref{pn}). Our polymer has $N=50$ monomers which translocate through a pore defined at the 
center of the membrane, taken as an $80 \times 80$ monoatomic lattice. The pore is created by the remotion of a single atom of 
the membrane.
\begin{figure}[tbph]
\includegraphics[width=11.88cm, height=9.24cm]{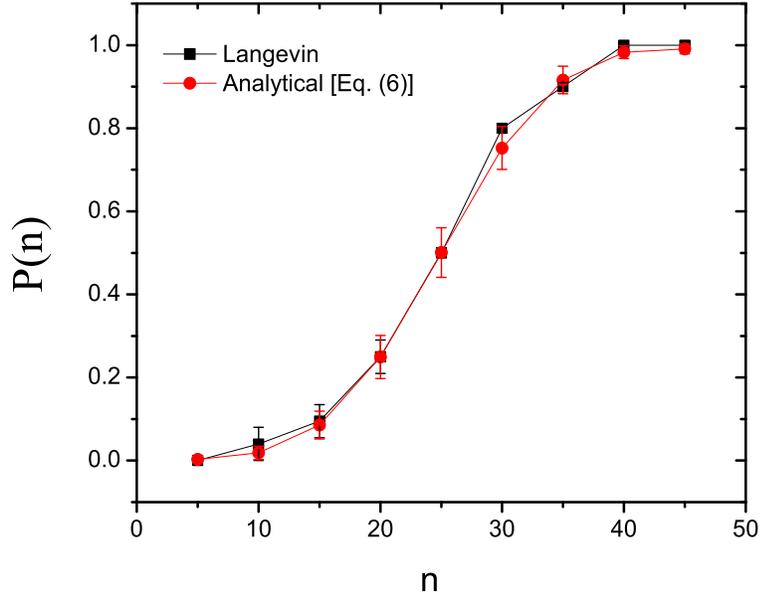}
\caption{The empirical (black squares) and analytical (red circles) probabilities (both denoted here by $P(n)$) of cis $\rightarrow$ trans 
complete translocations are compared. The variable $n$ stands for the initial number of monomers in the trans-side of the membrane. 
The analytical probabilities are computed from (\ref{prob}) and (\ref{pn}) with $\delta=0.88$ and $\nu=0.588$.}
\label{fig5}
\end{figure}
Following the usual Langevin modeling prescriptions \cite{huopa,wei}, the mononomer-monomer or monomer-membrane interactions are described by the Lenard-Jones potential
\be
U_{LJ}(r) = \left\{ \begin{array}{ll}
4 \epsilon [(\sigma/r)^{12}-(\sigma r)^6]+\epsilon  \ , \
 r \leq 2^{1/6} \sigma \\ 
0  \ , \   r > 2^{1/6} \sigma \ . \
\end{array}\right.
\ee
Consecutive monomers in the polymer chain interact, additionally, through the 
finitely extensible nonlinear elastic (FENE) potential,
\be
U_{F}(r) = - \frac{1}{2} k R_0^2 \ln [ 1 -(r/R_0)^2] \ . \
\ee
Monomers evolve, then, according to the Langevin equations,
\be
m \frac{d^2 \vec r_i}{dt} = - \sum_{ j \neq i} \vec \nabla_{r_i}  [U_{LJ}(r_{ij}) + U_F(r_{ij})] -\xi \frac{d \vec r_i }{dt} + \vec F_i(t) \ , \
\ee
where $r_{ij} = |\vec r_i - \vec r_j|$, $\xi$ is the dissipative constant and $\vec F_i(t)$ is the gaussian stochastic force which acts on 
the monomer with label $i$,
\bea
&&\langle \vec F_i(t) \rangle = 0 \ , \ \nonumber \\
&&\langle [\hat n \cdot \vec F_i(t)] [\hat n' \cdot \vec F_j(t') ] \rangle = 2 \hat n \cdot \hat n' k_B T \xi \delta_{ij} \delta(t-t') \ . \
\label{eq-motion}
\eea
Above, $\hat n$ and $\hat n'$ are arbitrary unit vectors, and $k_B$ and $T$ are the Boltzmann constant and the temperature, respectively.
By means of a suitable regularization of the stochastic force, we have implemented a fourth-order Runge-Kutta scheme for the
numerical simulation of (\ref{eq-motion}). Our simulation parameters are: $\epsilon=1.0$, $\sigma=1.0$ ($\sigma$ is also identified to the membrane lattice parameter), $\xi =0.7$, $k=7 \epsilon/\sigma^2$, $R_0 = 2 \sigma$, $k_B T = 1.2 \epsilon$. The simulation time step is taken as $3 \times 10^{-3}t_{LJ}$, where $t_{LJ} \equiv \sqrt{m \sigma^2 / \epsilon}$ is the usual Lenard-Jones time scale.

The initial configuration of the polymer has $n$ monomers on the trans-side of the membrane. Translocation is allowed to occur only after thermal equilibrium is reached for the cis and trans sectors of the polymer. The empirical probability of translocation to the cis-side of the membrane is then measured for $n=5,10,15,...,45$. For each $n$, we consider an ensemble of 70 complete translocations. 

In order to compare Eq. (\ref{pn}) with the empirical probability computed from the Langevin simulations, it is necessary to realize that 
the Markov approximation is likely to hold for the consecutive translocation of small monomer clusters, rather than for the consecutive translocation of individual monomers. We expect, thus, that even though there may be strong correlation effects between the translocation of consecutive monomers, these correlations become small from cluster to cluster (it is reasonable to assume that the cluster size is proportional to the polymer persistence length). We have found, as it is shown in Fig. 5, that a suggestive comparison between the empirical and analytical probabilities can be obtained considering clusters with the size of 5 monomers. More precisely, for a given value of $n$ (the initial number of trans-monomers) a red circle is plotted in Fig. 5 with coordinates $(n, P(n/5))$ where $P(n/5)$ is the splitting probability evaluated for a polymer which contains $N/5=10$ monomers.

The empirical error bars in Fig. 5 were evaluated by the partition of each translocation sample (which has 70 elements) into 7 sub-sets. The theoretical error bars, on the other hand, follow from elementary statistical considerations and are given by $\sqrt{P(n)(1-P(n))/70}$.

We note that due to slow variation of the translocation dynamic exponents with the polymer size, the existence of the small monomer ``markovian clusters", as introduced above, has a neglegible effect on the results established in Sec. III.  

\section{Conclusions}

We have discussed in this paper a general kinetic model of polymer translocation, within the framework of markovian stochastic process. In order to comply with the standard picture of translocation \cite{kardar}, we have put forward a Chapman-Kolomogorov equation with transition probabilities $p_n$ and $q_n$, which turn out to depend as power laws on the number of cis (or trans) monomers in the polymer chain. We have also established a closed analytical expression for the probability of complete polymer translocation in terms of arbitrary $p_n's$ and $q_n's$. The Chapman-Kolmogorov equation is then numerically solved for quasi-equilibrium regimes where the CKK picture of translocation is assumed to work. We have been able to find good agreement with the scaling results derived from previous realistic Langevin simulations \cite{wei}, only if a non-trivial scaling relation, due to finite size effects, is taken into account for the diffusion constant of the polymer center of mass, a phenomenon observed in simulations by Bhattacharya et al. \cite{bhatta}. Our results indicate, thus, that finite size effects, which have been so far left to a secondary role in most of the polymer translocation literature, should have fundamental importance in the analysis of real or numerical translocation experiments. It would be interesting, in particular, to revisit Langevin simulations of polymer translocation under various solvent qualities, having in mind the relevance of finite size corrections.

Subleading corrections to the predicted asymptotic scaling profiles are found to have a slow decay, a fact that could explain some of the controversy on the precise values of the scaling exponents, and the issue whether they are actually universal or not. The reasonable precision associated with not very large polymers in Langevin simulations (see the third row in Table I) could suggest at first that universality or the CKK picture needs revision. However, we regard this apparent difficulty as a peculiar effect due to subleading scaling exponents and to corrections on the polymer mobility exponent. We point out, in this respect, that the amplitude of subleading corrections can be sensitive to specific modeling details, an additional complication factor in the empirical evaluation of universal translocation scaling exponents.

The Markov chain modeling strategy can be used, in principle, as a valuable tool in further real or numerical translocation experiments. Once a set of transition probabilities $p_n$ and $q_n$ is determined in an experimental study of translocation, a comparison between the markovian expression (\ref{pn}) and the observed frequency of complete translocation as a function of the initial number of trans monomers can be carried out, in order to reveal (or not) the existence of memory effects in the translocation process. Actually, we have been able to perform a successfull test of Eq. (\ref{pn}) with the help of Langevin simulations and have noted that a re-interpretation of recent Langevin simulation results \cite{dubbeldam} suggests that polymer translocation diffuses in a normal way (as a linear function of time), which is in complete agreement with the Markov chain picture addressed here.

\vspace{0.2cm}

This work has been partially supported by CNPq and FAPERJ.

\end{document}